\documentclass[manuscript,screen]{acmart}

\usepackage{listings} 
\usepackage{multirow}

\usepackage[normalem]{ulem} 

\newcommand*{\RQTwo} [1] {And isn't this one even better ?}
\newcommand*{\RQThree} [1] {This one tops it all, doesn't it ?}

\definecolor{keywordcolor}{rgb}{0.0, 0.2, 0.6} 
\definecolor{commentcolor}{rgb}{0.3, 0.6, 0.3} 
\definecolor{stringcolor}{rgb}{0.2, 0.5, 0.2} 
\definecolor{classcolor}{rgb}{0.7, 0.5, 0.0} 
\definecolor{annotationcolor}{rgb}{0.6, 0.3, 0.0} 
\definecolor{numbercolor}{rgb}{0.4, 0.4, 0.4} 

\lstdefinestyle{javacode}{
  language=Java,
  basicstyle=\footnotesize\color{black}, 
  keywordstyle=\color{keywordcolor}, 
  commentstyle=\color{brown}\itshape, 
  stringstyle=\color{stringcolor}, 
  classoffset=1, 
  morekeywords={UploadImage}, 
  keywordstyle=[1]\color{classcolor}, 
  keywordstyle=[2]\color{annotationcolor}, 
  morekeywords=[2]{@Test}, 
  numbers=left,
  numberstyle=\tiny\color{numbercolor}, 
  stepnumber=1,
  frame=single,
  tabsize=4,
  showstringspaces=false,
  breaklines=true,
  captionpos=b,
  numbersep=10pt, 
  framexleftmargin=15pt, 
  xleftmargin=20pt 
}

\lstdefinelanguage{json}{
    basicstyle=\ttfamily\footnotesize\color{black}, 
    numbers=left, 
    numberstyle=\tiny\color{numbercolor}, 
    stepnumber=1, 
    numbersep=10pt, 
    showstringspaces=false, 
    breaklines=true, 
    frame=single, 
    tabsize=4, 
    framexleftmargin=15pt, 
    xleftmargin=20pt, 
    captionpos=b, 
    literate=
     *{0}{{{\color{blue}0}}}{1}%
      {1}{{{\color{blue}1}}}{1}%
      {2}{{{\color{blue}2}}}{1}%
      {3}{{{\color{blue}3}}}{1}%
      {4}{{{\color{blue}4}}}{1}%
      {5}{{{\color{blue}5}}}{1}%
      {6}{{{\color{blue}6}}}{1}%
      {7}{{{\color{blue}7}}}{1}%
      {8}{{{\color{blue}8}}}{1}%
      {9}{{{\color{blue}9}}}{1}%
      {:}{{{\bfseries\color{black}:}}}{1}%
      {,}{{{\bfseries\color{black},}}}{1}%
      {\{}{{{\color{black}\{}}}{1}%
      {\}}{{{\color{black}\}}}}{1}%
      {[}{{{\color{black}[}}}{1}%
      {]}{{{\color{black}]}}}{1}%
}

\AtBeginDocument{%
  }

\setcopyright{acmlicensed}
\copyrightyear{2025}
\acmYear{2025}
\acmDOI{XXXXXXX.XXXXXXX}
\acmJournal{TOSEM}
\acmVolume{0}
\acmNumber{0}
\acmArticle{0}
\acmMonth{0}
\acmISBN{978-1-4503-XXXX-X/2018/06}

\newcommand{\paperTitle}{Agentic LLMs for REST API Test Amplification: A Comparative Study Across Cloud Applications}

\begin{document}

\title{\paperTitle}

\author{Jarne Besjes}
\email{Jarne.Besjes@student.uantwerpen.be}
\orcid{0009-0001-3714-3940}
\affiliation{%
  \institution{University of Antwerp}
  \city{Antwerp}
  \state{Antwerp}
  \country{Belgium}
}

\author{Robbe Nooyens}
\email{Robbe.Nooyens@student.uantwerpen.be}
\orcid{0009-0003-4800-1651}
\affiliation{%
  \institution{University of Antwerp}
  \city{Antwerp}
  \state{Antwerp}
  \country{Belgium}
}

\author{Tolgahan Bardakci}
\authornote{Corresponding Author}
\email{tolgahan.bardakci@uantwerpen.be}
\orcid{0009-0007-1136-2065}
\affiliation{%
  \institution{University of Antwerp and Flanders Make}
  \city{Antwerp}
  \state{Antwerp}
  \country{Belgium}
}

\author{Mutlu Beyaz{\i}t}
\email{mutlu.beyazit@uantwerpen.be}
\orcid{0000-0003-2714-8155}
\affiliation{%
  \institution{University of Antwerp and Flanders Make}
  \city{Antwerp}
  \state{Antwerp}
  \country{Belgium}
}

\author{Serge Demeyer}
\email{serge.demeyer@uantwerpen.be}
\orcid{0000-0002-4463-2945}
\affiliation{%
  \institution{University of Antwerp and Flanders Make}
  \city{Antwerp}
  \state{Antwerp}
  \country{Belgium}
}

\renewcommand{\shortauthors}{Besjes et al.}

\begin{abstract}
Representational State Transfer (REST) APIs are a cornerstone of modern cloud-native systems. Ensuring their reliability demands automated test suites that exercise diverse and boundary-level behaviors. 
Nevertheless, designing such test cases remains a challenging and resource-intensive endeavor.
This study extends prior work on Large Language Model (LLM)–based test amplification by evaluating single-agent and multi-agent configurations across four additional cloud applications.
The amplified test suites maintain semantic validity with minimal human intervention.
The results demonstrate that agentic LLM systems can effectively generalize across heterogeneous API architectures, increasing endpoint and parameter coverage while revealing defects.
Moreover, a detailed analysis of computational cost, runtime, and energy consumption highlights trade-offs between accuracy, scalability, and efficiency.
These findings underscore the potential of LLM-driven test amplification to advance the automation and sustainability of REST API testing in complex cloud environments.
\end{abstract}

\ccsdesc[500]{Software and its engineering}
\ccsdesc[500]{Software and its engineering~Designing software}
\ccsdesc[500]{Software and its engineering~Development frameworks and environments}
\ccsdesc[500]{Computing methodologies~Artificial intelligence}

\keywords{Software Testing, Test Amplification, Large Language Models, Agentic Systems}

\received{31 October 2025}
\received[revised]{12 March XXX}
\received[accepted]{5 June XXXX}

\maketitle

\section{Introduction}
Representational State Transfer (REST) APIs are the backbone of modern cloud-native ecosystems, enabling scalable and loosely coupled communication among distributed services.
However, their flexibility comes with substantial testing challenges~\cite{verborgh2015fallacy}.
The vast number of possible input combinations, diverse data dependencies, and independently evolving service endpoints create a near-infinite interaction space, where only a fraction of inputs expose defects—a classic “needle in a haystack” scenario.
Ensuring the reliability of such APIs requires automated test suites that are both comprehensive and adaptive, capable of exploring boundary conditions and protocol corner cases that are often left untested.

Traditional automated test generation techniques have made progress toward this goal.
However, these approaches often lack contextual awareness of the existing test suite on the system under consideration.
In contrast, test amplification builds upon existing test cases, refining and extending them to increase coverage and fault detection while preserving the original structure and intent.
This approach better aligns with how developers write and maintain test cases in practice, and it has been shown to improve both code coverage and developer trust~\cite{test_amplification_ampyfier},~\cite{unit_test_amplification-meta}.
The recent integration of large language models (LLMs) into test amplification further extends this potential.
By reasoning about code semantics, documentation, and more contextual knowledge, LLMs can synthesize new test cases that are both meaningful and executable~\cite{schäfer2024empiricalevaluationusinglarge}.

In previous work, we introduced the first feasibility study on agentic LLM systems for REST API test amplification~\cite{nooyens2025agenticamplification}.
We compared single-agent and multi-agent architectures.
The results demonstrated that both approaches could autonomously amplify existing test suites, improving structural API coverage and exposing hidden defects in a benchmark cloud application.
The multi-agent configuration achieved higher coverage and bug detection rates but required more computational and energy resources, revealing a trade-off between performance and efficiency.
While these initial findings are promising, they are based on a single API, Swagger PetStore~\cite{SwaggerPetstore}, which limited their generalizability.
In this extended study, we address this limitation by applying our amplification framework to four additional real-world cloud applications with distinct architectures, endpoints, and complexity levels. 
We use the following cloud applications: Google Drive~\cite{google_drive_api}, Spotify~\cite{spotify_api}, VAmPI~\cite{vampi_api}, and Restful-Booker\cite{restfulbooker}.
Our goal is to evaluate whether LLM-driven test amplification generalizes across heterogeneous systems and to assess its consistency, scalability, and sustainability.
Through this large-scale evaluation, we make the following contributions.
\begin{enumerate}
    \item Cross-system validation: We extend LLM-based test amplification to multiple real-world REST APIs, demonstrating that both single-agent and multi-agent configurations generalize effectively beyond a single benchmark.
    \item Comparative analysis: We provide a comprehensive comparison of structural coverage, fault detection, and qualitative readability across different application domains.
    \item Efficiency assessment: We analyze computational cost, runtime, and energy consumption to reveal key trade-offs between amplification quality and resource efficiency.
    \item Practical insights: We discuss lessons learned regarding the effectiveness of the agentic systems and maintainability when integrating agentic LLMs into automated testing pipelines.
\end{enumerate}
The results confirm that agentic LLM systems provide a robust and generalizable solution for REST API test amplification.
The implemented approaches advance the state of automated software testing toward greater autonomy, scalability, and environmental awareness.

\section{Related Work}
Test amplification has been recognized as a promising approach for improving the effectiveness of automated test suites~\cite{test_amplification_ampyfier}.
Unlike test generation, which often delivers context-agnostic tests, test amplification leverages existing test logic to increase coverage, fault detection, and developer trust.
Prior work has demonstrated its benefits for unit testing, regression detection, and mutation coverage improvement, providing strong evidence of its practical relevance in industrial contexts~\cite{test_amplification_definition}~\cite{test_amplification_ampyfier}.

The recent integration of Large Language Models (LLMs) into software testing research has shifted attention toward more autonomous and adaptive test generation.
Several studies have shown that LLMs can synthesize unit test cases of high quality~\cite{yang2024evaluationlargelanguagemodels}~\cite{unit_test_amplification-meta}, repair failing test cases through feedback loops~\cite{schäfer2024empiricalevaluationusinglarge}, and even improve coverage through iterative reasoning~\cite{pizzorno2025coverupcoverageguidedllmbasedtest}.
These results demonstrate the potential of LLMs for test generation and test amplification.

LLM-based frameworks have been developed for REST API testing. 
APITestGenie~\cite{pereira2024apitestgenieautomatedapitest} and AutoRestTest~\cite{stennett2025autoresttesttoolautomatedrest} demonstrated that LLMs can generate executable REST API test cases successfully.
Beyond test generation, CoverUp~\cite{pizzorno2025coverupcoverageguidedllmbasedtest} and TELPA~\cite{yang2024enhancingllmbasedtestgeneration} introduced iterative and coverage-guided generation mechanisms that approach test amplification principles by refining generated test cases based on execution results.

Agentic and multi-agent frameworks, such as MetaGPT~\cite{hong2024metagptmetaprogrammingmultiagent} 
have shown that decomposing software engineering tasks into specialized roles can improve coordination and quality.
The research on AgentCoder~\cite{huang2024agentcodermultiagentbasedcodegeneration} indicated that separating the roles of programmer and test designer into multiple agents led to a 13\% to 15\% improvement in coverage compared to employing a single agent.
Pan et al. introduced CodeCoR~\cite{pan2025codecorllmbasedselfreflectivemultiagent}, a multi-agent framework that incorporates self-reflection to assess the performance of individual agents and their teamwork.
They found that CodeCoR outperforms existing baselines, achieving an average Pass@1 score of 77.8\%.
However, empirical evidence comparing single-agent and multi-agent LLM systems for software testing and test amplification remains limited.

Our prior feasibility study~\cite{nooyens2025agenticamplification} introduced agentic LLM systems for REST API test amplification, comparing single- and multi-agent configurations.
While that work established feasibility, its evaluation was limited to a single API, leaving questions about generalization and efficiency trade-offs unanswered.
This paper addresses that gap by extending the analysis to four additional cloud applications (Google Drive~\cite{google_drive_api}, Spotify~\cite{spotify_api}, VAmPI~\cite{vampi_api}, and Restful-Booker~\cite{restfulbooker}), providing new evidence on how agentic LLMs scale across domains and at what computational cost.

\section{Background}
\paragraph{REST APIs} are web interfaces that enable clients to retrieve and modify server resources through standard HTTP methods.
They follow REST principles, which promote stateless interactions, a resource-centric design, and a consistent interface in which URLs represent data entities~\cite{fielding2000rest}.

\paragraph{Test amplification} is a broad concept encompassing different activities that analyze and modify existing test suites, such as augmentation, optimization, enrichment, and refactoring~\cite{test_amplification_definition}.
Unlike test generation, test amplification focuses on creating new test cases based on existing ones rather than generating them from the ground up.
This strategy offers a key advantage, as the resulting amplified test cases naturally fit within the existing test structure and conventions.

\paragraph{Large Language Models (LLMs)} are advanced deep learning models trained on massive text to comprehend, generate, and reason about natural language~\cite{minaee2025largelanguagemodelssurvey}.
By leveraging transformer-based architectures, they demonstrate strong generalization capabilities across diverse language tasks, including program synthesis and code generation \cite{austin2021programsynthesislargelanguage}~\cite{rozière2024codellamaopenfoundation}.

\paragraph{Generative Pre-trained Transformer (GPT)} is a family of large language models developed by OpenAI~\cite{openai}, built upon the transformer architecture~\cite{zhao2025surveylargelanguagemodels}.
These models are trained to comprehend and produce human-like text, enabling natural dialogue and supporting a broad spectrum of language understanding and generation tasks.
Recent studies have demonstrated GPT's effectiveness across various software engineering domains, highlighting its potential to assist in code comprehension, synthesis, and automated testing~\cite{schäfer2024empiricalevaluationusinglarge}~\cite{Qi2024potentialofgpt}.

\paragraph{Agentic LLM systems} are centered on the concept of autonomous, goal-driven agents.
Such systems integrate large language models as decision-making components capable of perceiving tasks, taking actions, and adapting their behavior based on context and feedback~\cite{Wang_2024_surveyagents}.
They reason about problems, interact with tools or environments, and generate purposeful outputs aligned with predefined objectives.
Agentic systems generally fall into two categories: single-agent and multi-agent configurations.
In a single-agent configuration, a single LLM instance is responsible for all sub-tasks, while multi-agent systems (MAS) distribute responsibilities among specialized agents that collaborate toward a shared goal~\cite{bousetouane2025agenticsystemsguidetransforming}.

\section{Approaches}
\label{sec:approaches}
This section discusses the basics of two agentic LLM systems developed for amplification of REST API test cases \cite{nooyens2025agenticamplification}.

\subsection{Single-Agent Configuration}
Pereira et al. (2024) introduced APITestGenie~\cite{pereira2024apitestgenieautomatedapitest}.
They demonstrated that a single-agent LLM architecture can autonomously generate executable REST API test scripts with minimal manual effort, achieving success rates of up to 80\%.
Since APITestGenie focuses on test generation and has operational requirements distinct from amplification, we developed our own test amplification framework using a single-agent approach.
We adopt this architecture as our baseline given its conceptual simplicity, ease of implementation, and demonstrated effectiveness in prior studies.

\begin{figure}[ht]
    \centering
    \includegraphics[width=0.4\textwidth]{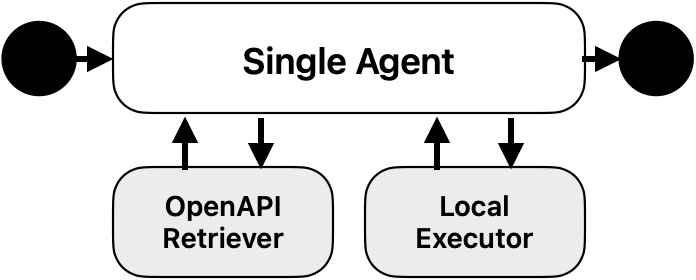}
    \caption{Single-Agent Configuration}
    \Description{Single-Agent Configuration Representative Figure}
    \label{fig:single-agent}
\end{figure}

Both systems are implemented using the LangGraph library~\cite{LangGraph}, an extension of the open-source LangChain framework designed for constructing stateful, agentic applications with explicit workflow control.
In the single-agent configuration, we employ LangGraph’s built-in ReAct (Reasoning and Acting) agent, which integrates task reasoning with decision-making about when to invoke external tools~\cite{yao2023reactsynergizingreasoningacting}.
The agent interacts with two specialized tools, as illustrated in \autoref{fig:single-agent}.
The first, the OpenAPI Retriever, enables the agent to query information from the API’s OpenAPI specification, such as endpoint definitions, available operations, parameters, request and response types, and status codes.
It can also retrieve schema details of specific data structures, allowing the agent to infer the API’s functional requirements and behavioral constraints.
The second tool, the Local Executor, stores generated test cases in a local repository and executes them, capturing console output that reveals passing and failing test cases, as well as syntax or runtime errors.

The effectiveness of an agent largely depends on the quality of its instructions.
Prior research has shown that prompt design plays a critical role in shaping LLM performance, with prompt structure and content directly influencing reasoning quality and output reliability~\cite{yang2024evaluationlargelanguagemodels}.
These studies emphasize the value of preliminary experimentation to identify optimal phrasing and the benefits of eliminating redundant or irrelevant information.
Following these principles, we crafted a prompt that delivers comprehensive yet concise context—covering the testing framework, available functions, and relevant local files—while explicitly instructing the agent to maximize REST API coverage across multiple criteria~\cite{test_coverage_criteria} and to explore edge cases likely to expose defects.
The final prompt, refined through iterative experimentation for both completeness and efficiency, is provided in the appendix Section~A.

\subsection{Multi-Agent Configuration}
Although a single-agent system serves as a strong baseline, prior work on agentic architectures indicates that complex tasks can benefit from decomposing the problem into smaller, specialized sub-tasks handled by distinct agents~\cite{hong2024metagptmetaprogrammingmultiagent}.
Each agent, guided by a focused, well-structured prompt, can efficiently manage its specific role in the test generation workflow.
We hypothesize that distributing responsibilities across multiple specialized agents yields more effective and coherent results than relying on a single agent to process a large, multifaceted prompt~\cite{yang2024evaluationlargelanguagemodels}.

\begin{figure}[htbp]
    \centering
    \includegraphics[width=0.9\textwidth]{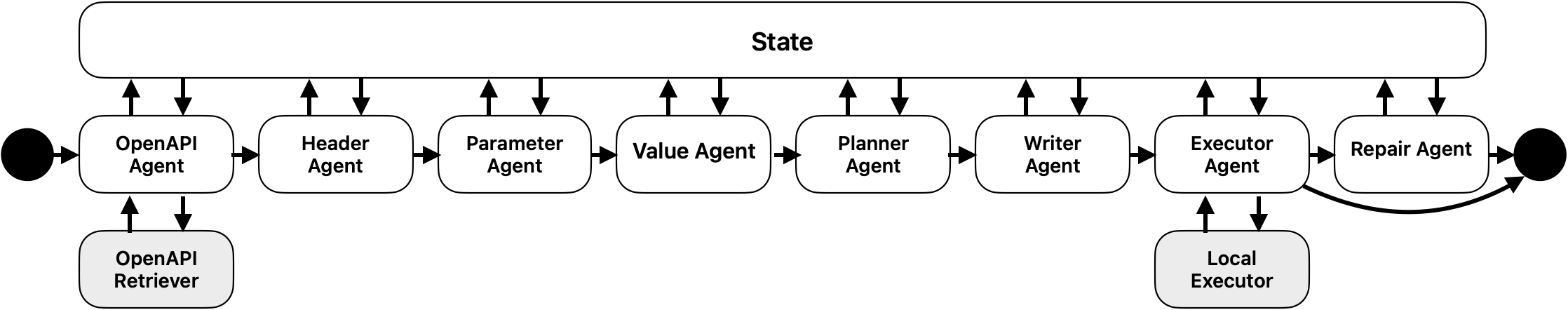}
    \caption{Multi-Agent Configuration}
    \Description{Multi-Agent Configuration Representative Figure}
    \label{fig:multi-agent-chain}
\end{figure}

Our multi-agent architecture, illustrated in \autoref{fig:multi-agent-chain}, was developed through a recursive decomposition of the overarching test suite amplification task.
This process yielded a sequential two-phase pipeline comprising test case planning and test code generation.
Such a separation of concerns—first determining what to test, then generating the corresponding code—follows an established principle in multi-agent software engineering, where high-level design and implementation are managed by distinct specialized agents~\cite{hong2024metagptmetaprogrammingmultiagent},~\cite{huang2024agentcodermultiagentbasedcodegeneration}.

\paragraph{The planning phase} is further divided into a set of specialized agent roles.
To ensure that each agent operates with sufficient contextual knowledge, the workflow begins with the OpenAPI Agent, responsible for retrieving relevant sections from the API’s OpenAPI specification.
This contextual information is then passed to a group of agents inspired by the AutoRestTest approach~\cite{stennett2025autoresttesttoolautomatedrest}: the Header Agent, Parameter Agent, and Value Agent.
These agents are prompted to propose input values targeting specific REST API coverage criteria~\cite{test_coverage_criteria}, such as triggering particular status codes or exploring parameter boundary conditions.
Their outputs are consolidated by the Planner Agent, which synthesizes the individual suggestions into a comprehensive list of test case descriptions.
For large and detailed specifications, the provided context can be substantial.
Because GPT models update their internal state sequentially from left to right~\cite{zhao2021calibrateuseimprovingfewshot}, the full contextual information is presented before the instruction segment, enabling the model to construct an accurate internal representation of the task.
Additionally, each agent employs Chain-of-Thought reasoning~\cite{wei2023chainofthoughtpromptingelicitsreasoning}, a prompting strategy that encourages step-by-step inference.
This technique enables agents to reason about complex relationships between parameters and values, often uncovering non-trivial combinations that expose unique API behaviors.
The complete input prompts are included in the appendix Section~A.

\paragraph{The generation phase} transforms the planned test cases into executable code.
This phase draws inspiration from feedback-driven frameworks such as CodeCoR~\cite{pan2025codecorllmbasedselfreflectivemultiagent}, which employs a Test Repair Agent to iteratively refine code based on execution feedback.
Following a similar principle, our architecture introduces a Test Writer Agent.
The Test Writer Agent is responsible for converting the test case descriptions produced by the Planner into executable test files.
The resulting code is then executed by a Test Executor Agent, which validates its correctness.
If compilation or runtime errors occur, the feedback is routed to a dedicated Test Repair Agent, which is prompted to identify and correct the issues in the original code.
All corresponding input prompts are available in the appendix Section~A.

\paragraph{The workflow} specifies the sequence of interactions among agents and the conditions governing transitions between agents and tools, orchestrated through the LangGraph framework~\cite{LangGraph}.
For instance, when the OpenAPI Agent requires access to the API specification, control transitions to a tool node that executes the OpenAPI Retriever and returns its output to the agent.
Similarly, the workflow includes a conditional branch to the Test Repair Agent, which is activated only when the Test Executor Agent identifies compilation or runtime errors.
This structured multi-agent orchestration enables a systematic approach to REST API test amplification, with each agent contributing domain-specific expertise to a particular stage of the testing process.
The workflow operates as a linear pipeline without feedback from execution results to the planning phase.
This design choice ensures deterministic evaluation and simplifies the experimental setup for our comparative study.
Although integrating an automated coverage analysis loop—such as direct feedback from the coverage tool was not feasible due to manual processing requirements, the modular architecture implemented in LangGraph readily supports future extensions.
Such extensions could enable iterative refinement, where coverage reports from the Test Executor Agent are automatically fed back to the Planner Agent to guide the generation of subsequent test cases.

\section{Evaluation}
\label{sec:evaluation}
The goal of our evaluation is to compare the performance of agentic LLM systems for REST API test amplification (see Section \ref{sec:approaches}) across multiple cloud applications of different size and complexity.
To this end, we investigate the following research questions comparatively.

\newcommand*{\RQTestStats} {\textit{Which portion of the amplified test cases are successful (and which portion fails)?}}
\newcommand*{\RQFailTypes} {\textit{What types of failures occur among the failed test cases?}}
\newcommand*{\RQCoverage} {\textit{What is the REST API  coverage achieved by the amplified test cases?}}
\newcommand*{\RQCosts} {\textit{What are the costs associated with LLM usage in terms of time, tokens, money, and energy consumption?}}
\newcommand*{\RQReadMaintain} {\textit{How are the readability and maintainability of the amplified test cases?}}
\begin{itemize}
    \item RQ1: \RQTestStats
    \item RQ2: \RQFailTypes
    \item RQ3: \RQCoverage
    \item RQ4: \RQCosts
    \item RQ5: \RQReadMaintain
\end{itemize}

\subsection{Systems Under Consideration}
To evaluate the generalizability of our approach, we applied both single-agent and multi-agent workflows to five different REST APIs.
Two of these systems—Google Drive~\cite{google_drive_api} and Spotify~\cite{spotify_api}—represent large-scale, production-grade cloud services.
The other three, the PetStore~\cite{SwaggerPetstore}, VAmPI~\cite{vampi_api}, and Restful-Booker~\cite{restfulbooker}, are locally hosted APIs deployed in controlled environments via Docker~\cite{docker}.
Together, these systems provide a balanced mix of complexity, domain diversity, and accessibility for reproducible experimentation.
~\autoref{tab:system_statistics} gives some descriptive statistics on these systems.

Real-world systems, such as Google Drive and Spotify, use multiple microservices for their operations.
Microservice architectures decompose the system into small, autonomous services that communicate through lightweight protocols such as HTTP or REST.
Each service is independently deployable and maintainable, enabling greater scalability and flexibility at the cost of increased system complexity~\cite{blinowski2022_monolithic_and_microservice}.

In contrast, open-source applications such as PetStore, VAmPI, and Resful-Booker use a monolithic architecture. 
Monolithic architectures organize an application as a single, unified executable where all components share the same process, database, and deployment lifecycle.
All business logic, data management, and user interface elements are tightly integrated within a single codebase, which simplifies development and deployment but limits modularity and scalability~\cite{blinowski2022_monolithic_and_microservice}.

\begin{table}[htbp]
  \centering
  \caption{Descriptive Statistics for the Systems used for evaluation}
  \Description{Descriptive statistics on the systems.}
  \label{tab:system_statistics}
  \begin{tabular}{lccccc}
    \toprule
                   & Domain                                 & Endpoints  & Data Models & System Architecture    \\
    \midrule
    Spotify        & Media / Music Streaming                & 89          & 93         & Multiple Microservices \\
    \midrule
    Google Drive   & Cloud Storage                          & 41          & 22         & Multiple Microservices \\
    \midrule
    PetStore       & E-commerce / Sample API                & 20          & 6          & Monolithic             \\
    \midrule
    VAmPI          & Security / Vulnerable API              & 14          & 0          & Monolithic             \\
    \midrule
    Restful-Booker & Booking / Hotel Reservation            & 8           & 5          & Monolithic             \\
    \bottomrule
  \end{tabular}
\end{table}

\paragraph{Spotify} The Spotify Web API~\cite{spotify_api} offers programmatic access to Spotify’s music streaming platform.
It exposes endpoints for managing playlists, retrieving track metadata, and interacting with user libraries.
As a cloud-native service with authentication and rate-limiting constraints, it provides a realistic scenario for assessing how well LLM-based testing agents adapt to authenticated and stateful APIs.

\paragraph{Google Drive} The Google Drive API~\cite{google_drive_api} is a cloud-based service that enables developers to programmatically manage files and folders in Google Drive.
It provides a comprehensive REST interface supporting file uploads, metadata queries, sharing configurations, and permission management.
Its rich functionality and strict authentication requirements make it an ideal benchmark for evaluating the scalability and robustness of REST API test amplification in real-world cloud ecosystems.

\paragraph{PetStore}
The PetStore API~\cite{SwaggerPetstore} is an example cloud application provided by the OpenAPI Initiative for demonstrating the structure and semantics of RESTful API specifications.
It models an e-commerce application that manages entities such as pets, users, and store orders.
Although not a real-world system, it was widely used as a standard benchmark, enabling straightforward reproducibility and comparability across studies~\cite{dias2025_PetStore_ref}~\cite{saha2025_PetStore_ref}.

\paragraph{VAmPI} VAmPI (Vulnerable RESTful API)~\cite{vampi_api} is an intentionally insecure REST API designed for security testing and vulnerability research.
It includes multiple OWASP (Open Worldwide Application Security Project)~\cite{owasp} top 10 vulnerabilities.
We deployed VAmPI locally in a Docker environment to evaluate whether the amplified test cases can reveal defects and exercise edge cases.

\paragraph{Restful-Booker}
Restful-Booker~\cite{restfulbooker} is a lightweight web service designed specifically for testing REST API concepts.
It exposes a set of endpoints supporting core HTTP operations such as \texttt{GET}, \texttt{POST}, \texttt{PUT}, \texttt{PATCH}, and \texttt{DELETE}.
The API provides a simple booking domain model that includes entities such as \texttt{Booking}, \texttt{BookingDates}, and \texttt{Auth}.
We deployed Restful-Booker locally using Docker~\cite{docker} to create a reproducible and isolated environment for evaluation.

Collectively, these five systems capture a diverse spectrum of REST API characteristics.
They range from the ones that contain multiple microservices (Spotify and Google Drive) to lightweight, monolithic APIs (PetStore, VAmPI, and Restful-Booker).
This diversity enables a meaningful comparison of our single- and multi-agent setups across various system architectures.

By combining both production-grade and controlled environments, our evaluation provides balanced evidence of the approach’s generalization and practical applicability to real-world REST API testing scenarios.

\subsection{Approach Parameters}
The following parameters are applied consistently across all experimental configurations, similar to what was done in our previous work~\cite{nooyens2025agenticamplification}.
\begin{itemize}
    \item We employ OpenAI’s GPT-4o mini model as the underlying LLM, selected for its balance between output quality and computational cost.
    \item The temperature parameter controls the stochasticity of the model: values near zero yield deterministic outputs, while higher values introduce greater variability.
    To ensure reproducibility and adherence to the OpenAPI specification, we set the temperature to 0, minimizing randomness in generated responses.
    \item Each system is initialized with a baseline happy-path test case, treated as the input test suite, along with function definitions for the framework’s \texttt{GET}, \texttt{POST}, \texttt{DELETE}, \texttt{PATCH}, and \texttt{PUT} operations.
\end{itemize}
All materials, including parameter configurations, are publicly available in our reproduction package\footnote{\url{https://figshare.com/s/0b1dde383f1e73984b97}}.

\subsection{Evaluation Criteria}
\label{sec:evaluationcriteria}
To address our research question, we assess the systems using various metrics, grounded in established practices from REST API testing literature.

For RQ1 and RQ2, we collect test execution statistics capturing the systems’ pass/fail outcomes.
For RQ3, we evaluate structural API coverage to quantify the extent of the tested functionality from a black-box perspective~\cite{restats}~\cite{Kim_2022}.
For RQ4, we analyze LLM usage statistics to assess the practical cost and computational efficiency of employing generative models~\cite{pereira2024apitestgenieautomatedapitest}.
Finally, for RQ5, we examine code readability, reflecting the qualitative and maintainability aspects of the generated test cases~\cite{buse2010learningametricforcodereadability}~\cite{schäfer2024empiricalevaluationusinglarge}, a dimension in which large language models often excel.

\textit{Test execution statistics} capture the core outcomes observed from the execution of the amplified test cases.
We report the total number of generated test cases, as well as the counts of successful and failed cases.
The classification of each test case was automatically determined using PyCharm~\cite{jetbrains_pycharm} and a custom script that extracted information from failed test cases.

Before we collect text execution statistics, we address trivial issues and make minor refinements to the amplified code, including:
\begin{itemize}
    \item Placeholders that are missing actual system-specific values are assigned correct values.
    \item Missing import statements are added.
    \item Parameters of \texttt{get}, \texttt{post}, \texttt{put}, \texttt{delete} and \texttt{patch} methods are fixed.
\end{itemize}

The refinement process requires no more than one hour in total.
The duration was measured manually using a clock by the first author.
We report the time required for the refinement in the section~\ref{sec:results}, Table~\ref{tab:test_statistics}.

\textit{Structural API coverage} metrics are computed across several dimensions, including path, operation, status class, status code, response type, request type, and parameter coverage~\cite{test_coverage_criteria}.
Restats~\cite{restats} is a tool for computing coverage values for these criteria. 
It is also used in previous studies~\cite{bardakci2025testamplificationrestapis}~\cite{nooyens2025agenticamplification}. 
However, in our experiments, we encountered several issues about Restats that prevented the successful computation of the results, including but not limited to the following:
\begin{itemize}
    \item \textit{Limited compatibility with different OpenAPI specification formats.}
    As we expanded our evaluation to include additional cloud applications, we quickly discovered that the tool failed to handle certain specification styles and versions (e.g., OpenAPI 2.0 vs 3.0) correctly.
    \item \textit{Limited matching requirements when processing execution logs.}
    Despite correctly testing specific parameters, the tool often made calculation errors due to strict exact-match requirements during log parsing.
\end{itemize}

Consequently, we developed and used our own tool, RESTCov~\cite{RESTCov_2025}, to calculate structural REST API coverage in this study.
RESTCov provides a more reliable and flexible mechanism for quantifying API-level test coverage.

\textit{LLM usage statistics} encompasses measurements of token usage, running time of the LLM setups, monetary cost, and energy consumption.
The monetary cost is estimated by multiplying the total number of input and output tokens by OpenAI’s published pricing rate.
All statistics, except the LLM setup running times, are collected using LangSmith~\cite{langsmith2023}, an open-source framework for tracking LLM execution metrics.
The running time of the LLM setups is calculated by Jupyter Interactive Notebook~\cite{jupyter_notebook} for the end of each agentic setup's successful run.
In addition to these conventional measures, we also assess energy consumption.
Given the increasing computational footprint of large-scale AI models, quantifying energy usage is essential for evaluating the sustainability of LLM-driven approaches.
Considering the smaller parameter size of GPT-4o-mini (estimated to be around 8 billion parameters compared to GPT-4o’s 200 billion parameters)~\cite{abacha2025medecbenchmarkmedicalerror_parametercounts}, 
a typical GPT-4o query consumes approximately 0.3~watt-hours per 500 tokens or about 0.00006 watt-hours per token~\cite{epoch2024chatgptenergy}.

\textit{Readability} is a key non-functional quality attribute in software engineering, closely linked to code comprehension and maintainability~\cite{buse2010learningametricforcodereadability}.
To assess readability, we evaluate the generated test cases using a structured checklist comprising the following criteria:
\begin{itemize}
    \item \textit{Meaningful naming.} Whether the generated test methods and variable names are descriptive and conform to the naming conventions of the target programming language.
    \item \textit{Structural coherence:} Whether the generated test cases maintain stylistic consistency with the initial example test provided as a template.
    \item \textit{Idiomatic correctness.} Whether the code appropriately applies standard constructs of the target language and testing framework.
\end{itemize}
The readability assessment follows a two-step review process: one author conducted an initial evaluation, and a second author independently verified the results.
Given the qualitative nature of this analysis, representative examples of generated test cases are included in Appendix, Section~\ref{sec:testcases}.

\section{Results And Discussion}
\label{sec:results}
Here, we answer the research questions RQ1-RQ5 presented in Section \ref{sec:evaluation}.
While the amplified tests for PetStore are taken from our previous study~\cite{nooyens2025agenticamplification}, we recalculate the coverage using our tool~\cite{RESTCov_2025} and include recently computed results for the new systems, namely Spotify, Google Drive, VAmPI, and Restful-Booker.

\subsection{RQ1. \RQTestStats}
Our test execution statistics are given in~\autoref{tab:test_statistics}.
As mentioned in Section~\ref{sec:evaluationcriteria}, these statistics were collected after minor refinements.
The refinement effort seems to be increasing depending on the complexity of the system.
Our most complex system, Spotify, has ~\textit{55-single-agent}, and ~\textit{50-multi-agent} minutes of effort required.
The Google Drive follows up by ~\textit{5,05-single-agent}, and ~\textit{18,48-multi-agent} minutes.
However, when system complexity decreases, the effort is none to around a minute.

\autoref{tab:test_statistics} shows that the number of failed test cases is significantly greater than the number of successful test cases. 
The portions of successful test cases range from $\%16.07$ to $\%49.15$ for the single-agent configuration, and from $\%6.06$ to $\%46.21$ for the multi-agent configuration. 
These results suggest that as the system gets more complex, the portion of successful test cases becomes smaller.
In contrast, the number of failed cases increases in complex systems, indicating that LLMs struggle in more difficult contexts.

On the other hand, when we compare single-agent and multi-agent configurations, the portions of successful test cases are within $3\%$ of each other, except for VAmPI, where this ratio drops by $19.50\%$ when a multi-agent configuration is used.
Thus, in general, single-agent and multi-agent configurations yield similar portions of successful test cases.

When we look at the number of test cases generated by these configurations, the multi-agent configuration results in $\%8.16$ to $\%22.88$ greater numbers of generated test cases and in $\%10.44$ to $\%47.17$ greater numbers of failed test cases than the single-agent configuration in all the systems. 
For successful test cases, the results are slightly different. 
Multi-agent configuration results in $\%15.52$ to $\%34.34$ greater numbers of successful test cases in four systems (Pet Store, Restful-Booker, Google Drive, and Spotify), and $\%37.78$ fewer number of successful test cases in one system (VAmPI).

\begin{table}[htbp]
  \centering
  \caption{Test Execution Statistics}
  \Description{Test results after the execution.}
  \label{tab:test_statistics}
  \begin{tabular}{llcccc}
    \toprule
                                    &                            & Single-Agent & Multi-Agent \\ 
    \midrule
    \multirow{4}{*}{Spotify}        & Generated Test Cases       & 616          & 704         \\
                                    & Successful Test Cases      & 100          & 112         \\
                                    & Failed Test Cases          & 516          & 592         \\
                                    & Refinement Time (m)        & 55           & 50          \\
    \midrule                             
    \multirow{4}{*}{Google Drive}   & Generated Test Cases       & 189          & 245         \\
                                    & Successful Test Cases      & 89           & 46          \\
                                    & Failed Test Cases          & 100          & 199         \\
                                    & Refinement Time (m)        & 5,05         & 18,48       \\
    \midrule
    \multirow{4}{*}{PetStore}       & Generated Test Cases       & 118          & 145         \\
                                    & Successful Test Cases      & 58           & 67          \\
                                    & Failed Test Cases          & 60           & 78          \\
                                    & Refinement Time (m)        & N/A          & N/A         \\
    \midrule
    \multirow{4}{*}{VAmPI}          & Generated Test Cases       & 98           & 106         \\
                                    & Successful Test Cases      & 45           & 28          \\
                                    & Failed Test Cases          & 53           & 78          \\
                                    & Refinement Time (m)        & 1            & 0           \\
    \midrule
    \multirow{4}{*}{Restful-Booker} & Generated Test Cases       & 34           & 41          \\
                                    & Successful Test Cases      & 10           & 12          \\
                                    & Failed Test Cases          & 24           & 29          \\
                                    & Refinement Time (m)        & 0,5          & 0           \\
    \bottomrule
  \end{tabular}
\end{table}

\subsection{RQ2. \RQFailTypes}
\autoref{tab:failedtestcategories} categorizes failed test cases based on an automated process into two: test cases failing with assertion errors, and test cases failing with other runtime errors.

\begin{table}[htbp]
  \centering
  \caption{Categorization of Failed Test Cases}
  \Description{Details about the failed test cases, categorized by different labels.}
  \label{tab:failedtestcategories}
  \begin{tabular}{llcc}
    \toprule
                                    & Metric                       & Single-Agent  & Multi-Agent  \\
    \midrule
    \multirow{3}{*}{Spotify}        & Failed Test Cases            & 516            & 592         \\
                                    & Assertion Errors             & 509            & 538         \\
                                    & Other Run-time Errors        & 7              & 54          \\
    \midrule
    \multirow{3}{*}{Google Drive}   & Failed Test Cases            & 100            & 199         \\
                                    & Assertion Errors             & 100            & 183         \\
                                    & Other Run-time Errors        & 0              & 16          \\
    \midrule
    \multirow{3}{*}{PetStore} 
                                    & Failed Test Cases            & 60             & 78          \\
                                    & Assertion Errors             & 52             & 74          \\
                                    & Other Run-time Errors        & 8              & 4           \\
    \midrule
    \multirow{3}{*}{VAmPI}          & Failed Test Cases            & 53             & 78          \\
                                    & Assertion Errors             & 51             & 76          \\
                                    & Other Run-time Errors        & 2              & 2           \\
    \midrule
    \multirow{3}{*}{Restful-Booker} & Failed Test Cases            & 24             & 29          \\
                                    & Assertion Errors             & 24             & 23          \\
                                    & Other Run-time Errors        & 0              & 6           \\
    \bottomrule
  \end{tabular}
\end{table}

For both single-agent and multi-agent configurations, the majority of failures across all systems stem from assertion errors.
It indicates that most of the generated tests reached the expected execution path but produced mismatched outcomes.
The percentage of test cases failing with assertion errors ranges from $86.67\%$ to $100\%$ for the single-agent configuration and from $79.31\%$ to $97.44\%$ for the multi-agent configuration.
This suggests that the generated inputs and execution contexts are, in general, syntactically correct, yet they produce outcomes that are either semantically invalid or reveal genuine defects in the system.

Across different configurations, the multi-agent configuration generally results in a smaller portion of test cases failing with assertion errors.
For Spotify, Google Drive, and Restful-Booker, these portions are, respectively, $7.87\%$, $8.04\%$, and $20.69\%$ smaller; whereas, for VAmPI and PetStore, they are, respectively, $1.26\%$ and $9.47\%$ greater.
The results indicate that, for relatively more complex systems, the multi-agent configuration causes an increase in the proportion of test cases failing with other runtime errors (or a decrease in the proportion failing with assertion errors) compared to the single-agent configuration.

\subsection{RQ3. \RQCoverage}
Structural API coverage achieved across different configurations and systems is shown in~\autoref{tab:structural_API_coverage}.

In both single-agent and multi-agent configurations, the amplified test suites result in significant increases across all coverage criteria.
They achieve near-complete path and operation coverages ($\geq 93\%$) across all systems.
Both configurations demonstrate that the amplification process successfully yields test cases that exercise nearly all documented API endpoints.
For the single-agent configuration, coverage remains consistently high across simpler systems such as Restful-Booker and VAmPI. 
It is slightly lower in some dimensions, such as status and status class, which depend on diverse response codes.
In more complex systems like Spotify and Google Drive, the coverage of these response-related dimensions decreases substantially, reflecting the difficulty of reaching certain execution parts.
In the multi-agent configuration, the range of coverage values broadens.
The systems with simpler structures (Petstore, VAmPI, and Restful-Booker) maintain complete or near-complete coverage across most dimensions, whereas the most complex systems (Spotify and Google Drive) exhibit lower coverage in some categories.
For example, status coverage on Google Drive and Spotify, and response type coverage on Google Drive, remains relatively lower.

\begin{table}[htbp]
  \centering
  \caption{Structural API Coverage}
  \Description{The structural API coverage of each cloud application.}
  \label{tab:structural_API_coverage}
  \begin{tabular}{llccc}
    \toprule
                                      &               &Initial Coverage & Single-Agent  & Multi-Agent \\
    \midrule
    \multirow{7}{*}{Spotify}          & Path          & 1,5\%           & 98,5\%        & 100,0\%     \\
                                      & Operation     & 1,1\%           & 93,3\%        & 94,4\%      \\
                                      & Status Class  & 0,6\%           & 56,2\%        & 59,0\%      \\
                                      & Status        & 0,3\%           & 31,1\%        & 28,9\%      \\
                                      & Response Type & 1,1\%           & 92,1\%        & 94,4\%      \\
                                      & Request Type  & 0,0\%           & 0\%           & 47,4\%      \\
                                      & Parameter     & 0,5\%           & 71,8\%        & 92,1\%      \\
    \midrule
    \multirow{7}{*}{Google Drive}     & Path          & 4,2\%           & 100,0\%       & 100,0\%     \\
                                      & Operation     & 2,4\%           & 100,0\%       & 97,6\%      \\
                                      & Status Class  & 2,4\%           & 31,7\%        & 9,8\%       \\
                                      & Status        & 2,4\%           & 31,7\%        & 9,8\%       \\
                                      & Response Type & 4,1\%           & 68,8\%        & 18,8\%      \\
                                      & Request Type  & 0,0\%           & 86,7\%        & 93,3\%      \\
                                      & Parameter     & 0,6\%           & 36,2\%        & 53,2\%      \\
    \midrule
    \multirow{7}{*}{PetStore}         
                                      & Path          & 7\%              & 100\%        & 100\%       \\
                                      & Operation     & 5\%              & 95\%         & 100\%       \\
                                      & Status Class  & 4\%              & 56,5\%       & 63,8\%      \\
                                      & Status        & 3\%              & 36,1\%       & 37\%        \\
                                      & Response Type & 3\%              & 50,9\%       & 87,7\%      \\
                                      & Request Type  & 9\%              & 72,2\%       & 93,9\%      \\
                                      & Parameter     & 11\%             & 84\%         & 82\%        \\                                 
    \midrule
    \multirow{7}{*}{VAmPI}            & Path          & 16,7\%          & 100,0\%       & 100,0\%     \\
                                      & Operation     & 14,3\%          & 100,0\%       & 100,0\%     \\
                                      & Status Class  & 8,7\%           & 73,9\%        & 65,2\%      \\
                                      & Status        & 7,1\%           & 60,7\%        & 64,3\%      \\
                                      & Response Type & 14,3\%          & 100,0\%       & 100,0\%     \\
                                      & Request Type  & 0,0\%           & 100,0\%       & 100,0\%     \\
                                      & Parameter     & 0,0\%           & 100,0\%       & 100,0\%     \\
    \midrule
    \multirow{7}{*}{Restful-Booker}   & Path          & 25,0\%          & 100\%         & 100\%       \\
                                      & Operation     & 12,5\%          & 100\%         & 75\%        \\
                                      & Status Class  & 6,7\%           & 53,3\%        & 33\%        \\
                                      & Status        & 6,7\%           & 33,3\%        & 26,7\%      \\
                                      & Response Type & 12,5\%          & 37,5\%        & 12,5\%      \\
                                      & Request Type  & 0,0\%           & 100\%         & 100\%       \\
                                      & Parameter     & 6,7\%           & 100\%         & 60\%        \\
    \bottomrule
  \end{tabular}
\end{table}

When comparing single-agent against multi-agent configurations, the multi-agent configuration generally maintains or slightly improves coverage relative to the single-agent configuration. However, that is not in all cases.
This improvement is particularly evident in categories that require exploring diverse request structures, such as request type and parameter coverage.
Except for Restful-Booker, the multi-agent configuration achieves higher coverage across all systems in these categories.
For example, in Google Drive and PetStore, the multi-agent configuration achieves higher parameter coverage.
However, for certain response-related metrics (e.g., status and response type), multi-agent coverage drops compared to the single-agent configuration, especially in Google Drive and Restful-Booker.
This may be attributed to agents exploring broader input combinations that lead to unhandled or repetitive responses, thereby reducing the diversity of observed status codes.

Overall, the results show that while both configurations achieve high structural coverage, the multi-agent approach excels at input-space exploration.
In contrast, in some occurrences, the single-agent configuration yields more stable response-space coverage.

These findings highlight that amplification through agentic configurations can enhance API exploration but introduces trade-offs in response diversity, particularly in complex or stateful systems.
The high endpoint and operation coverage across all systems demonstrates that the amplification pipeline effectively broadens structural reach.
Yet, differences in status-related metrics reveal that achieving comprehensive behavioral coverage still depends on system complexity and the agents’ ability to coordinate.

\subsection{\RQCosts}
LLM usage statistics are summarized in \autoref{tab:llmstatistics}.
These metrics offer insights into four key dimensions: execution time, token utilization, monetary costs, and estimated energy consumption.

\begin{table}[htbp]
  \centering
  \caption{LLM Usage Statistics}
  \Description{The results about how many resources have been used for agentic systems across all applications.}
  \label{tab:llmstatistics}
  \begin{tabular}{llcc}
    \toprule
                                       & Metric              & Single-Agent        & Multi-Agent \\
    \midrule
    \multirow{4}{*}{Spotify}           & Time (m)            & 41,6                & 107,72      \\
                                       & Tokens              & 211.210              & 321.888    \\
                                       & Cost of Usage (\$)  & 0,04                & 0,09        \\
                                       & Energy Usage (Wh)   & 12,673              & 19,313      \\
    \midrule
    \multirow{4}{*}{Google Drive}      & Time (m)            & 17,65               & 66          \\
                                       & Tokens              & 225.424             & 524.096     \\
                                       & Cost of Usage (\$)  & 0,04                & 0,15        \\
                                       & Energy Usage (Wh)   & 13,525              & 31,446      \\
    \midrule
    \multirow{4}{*}{PetStore}          & Time (m)            & 7,52                & 22,45       \\
                                       & Tokens              & 71.186              & 162.355     \\
                                       & Cost of Usage (\$)  & 0,02                & 0,05        \\
                                       & Energy Usage (Wh)   & 4,27                & 9,74        \\
    \midrule
    \multirow{4}{*}{VAmPI}             & Time (m)            & 6,02                & 21,75       \\
                                       & Tokens              & 109.922             & 91.698      \\
                                       & Cost of Usage (\$)  & 0,02                & 0,03        \\
                                       & Energy Usage (Wh)   & 6,595               & 5,502       \\
    \midrule
    \multirow{4}{*}{Restful-Booker}    & Time (m)            & 2,02                & 6,87        \\
                                       & Tokens              & 30.832              & 51.071      \\
                                       & Cost of Usage (\$)  & 0,01                & 0,02        \\
                                       & Energy Usage (Wh)   & 1,850               & 3,064       \\
    \bottomrule
  \end{tabular}
\end{table}

In both configurations, resource consumption generally scales with the system's complexity. 
In the single-agent configuration, Spotify and Google Drive exhibit the highest LLM running times (41,6 and 17,65 minutes, respectively).
Additionally, the token usage, cost, and energy consumption increase with larger API systems.
Simpler systems such as PetStore, VAmPI, and Restful-Booker show significantly lower costs across all dimensions.
This indicates that the model’s reasoning and generation demands increase proportionally with API size and call diversity.
A similar pattern is observed in the multi-agent configuration, where execution time, token usage, and cost rise sharply for Google Drive and Spotify, while remaining moderate for smaller systems.
The only notable deviation occurs in VAmPI, where token and energy usage decrease slightly in the multi-agent configuration.
This occurs because input tokens are more expensive than output tokens, leading to a lower total cost when fewer inputs are processed.
Overall, system complexity strongly correlates with LLM resource utilization.

When comparing configurations, except for the VAmPI, token usage, the multi-agent configuration consistently consumes more resources than the single-agent one.
The increase ranges from roughly 2 times (Restful-Booker) to nearly 4 times (Google Drive). 
This increase is most evident in execution time and token counts, which also directly influence monetary cost and energy consumption.
The cost difference, while small in absolute terms (e.g., 0.04\$ and 0.15\$ for Google Drive), demonstrates the cumulative effect of agentic coordination.
Systems like VAmPI and Restful-Booker show relatively modest increases in cost and energy consumption, reflecting their limited interaction space and smaller API footprint.
The results reveal that the computational and token-level overhead introduced by multi-agent collaboration grows with the system’s size and the complexity of its interactions.

Overall, the results highlight a clear trade-off between exploration depth and resource efficiency.
Multi-agent amplification improves test diversity and API coverage (as seen in RQ3) but incurs higher temporal and energetic costs. 
For large systems such as Spotify and Google Drive, the multi-agent approach can be resource-intensive but beneficial for output diversity.
In contrast, for lightweight APIs (VAmPI, Restful-Booker), the single-agent configuration offers a balance between cost and test amplification quality.

\subsection{RQ5. \RQReadMaintain}
\textit{Readability} is a notable strength of LLMs in software development.
They excel at generating code that is natural and coherent.
They also demonstrate an understanding of underlying semantics~\cite{schäfer2024empiricalevaluationusinglarge}.
Using the readability checklist described in ~\ref{sec:evaluationcriteria}, we observed that the generated code consistently meets high-quality standards.
Given the qualitative assessment of readability criteria, we share some examples of amplified test cases in Section \ref{sec:testcases}.
In addition, to illustrate their effectiveness on readability, we analyze some of the amplified test cases in detail.

In a test case from Google Drive, given in Listing~\ref{lst:testGetDrivesWithMaxPageSize}, we observe that agentic configurations can successfully create boundary scenarios, which is critical for revealing bugs in software systems.
Although we only give an example for the~\textit{GET /files} endpoint, our amplified test case can exercise boundary testing for another endpoint and method, namely~\textit{GET /drives}.
While exercising another endpoint, it observes and preserves the overall structure of the framework~(\ref{lst:framework}), which significantly contributes to \textit{maintainability}.

In a test case example from Spotify, given in Listing~\ref{lst:developerTestCaseSpotify-I}, we observe that agentic configurations can successfully create additional happy-path scenarios based on what we provide them.
Although we only give an example for the~\textit {GET /artists} endpoint, our amplified test case can exercise the happy path for another endpoint and method, namely~\textit{GET /search}.
As we see in the test case, it follows Java naming conventions, and the names (variables, functions) are semantically meaningful.

Overall, our assessment shows that the amplified test cases maintain a high level of readability and structural consistency across different systems and agentic configurations.
For further assessment, we publicly share all materials, including the test code, in our reproduction package\footnote{\url{https://figshare.com/s/0b1dde383f1e73984b97}}.

\section{Threats to Validity}

Several factors may influence the outcomes of our approach and, consequently, affect the validity of our findings.
We discuss these threats in four categories.

\subsection{Construct Validity}
Our evaluation relies primarily on structural API coverage and the number of successful test cases as indicators of test amplification effectiveness.
Although these are widely accepted measures in REST API testing research~\cite{Kim_2022}, they have inherent limitations.
High coverage values do not necessarily imply that the most critical or complex API components have been thoroughly tested.
To address this concern, we developed RESTCov~\cite{RESTCov_2025} to accurately calculate structural API coverage and support a wider range of OpenAPI specifications and cloud applications.
Our previously used tool, Restats~\cite{restats}, as mentioned in Section~\ref{sec:evaluationcriteria}, had several limitations, and certain behavioral errors led to slightly inaccurate results.
Thus, when we recalculate the experiments from Nooyens et al.~\cite{nooyens2025agenticamplification} with RESTCov, we obtain more accurate coverage values than those previously calculated.
In addition to RESTCov, we use additional metrics to better capture the efficiency and trade-offs of agentic systems.

\subsection{Internal Validity}
A primary threat to internal validity concerns the LLMs’ prior knowledge of the systems under consideration.
This issue is particularly relevant for PetStore, VAmPI, and Restful-Booker, as these are open-source applications with extensive online documentation and publicly available examples.
To mitigate this threat, we built an automated system and evaluated it on two additional industrial-grade applications, Google Drive and Spotify.
This allows us to demonstrate and compare the designed system's ability to operate effectively across multiple cloud-based environments.
Including Spotify and Google Drive strengthens the comparison and demonstrates the effectiveness of the proposed approach beyond open-source benchmarks.
Future work can further minimize this threat by employing local LLMs or models trained on restricted corpora without access to the target systems’ source or test artifacts.

\subsection{External Validity}
Although the comparison provides valuable insights across both open-source and industrial-grade applications, the generalizability of the findings is still a question.
To increase the generalizability of our results, we use five cloud-based systems across different domains, complexities, and architectures.
One can always include more systems to confirm the consistency of the observed trends and better assess the external validity of the proposed approach.

\subsection{Reliability}
LLMs are non-deterministic and can generate different outputs for the same input.
This limits the reproducibility of the experiments.
To reduce this limitation, we use a fully automated system that ensures consistent execution and allows others to repeat or extend the experiments.

Another problem is that different language models—whether larger or smaller—may yield different outcomes.
Our results are generated using the GPT-4o-mini model provided by OpenAI, which contains an estimated 8 billion parameters, compared to the estimated 200 billion in GPT-4o~\cite{abacha2025medecbenchmarkmedicalerror_parametercounts}.
Different language models—whether larger or smaller—may yield different outcomes.

Also, GPT models are closed-source and version-controlled by OpenAI; potential future updates could influence model behavior, introducing variability in results or performance.

Readability is evaluated qualitatively by the authors.
Although we follow a structured review process to minimize subjectivity, individual judgments may differ across evaluators.
To support transparency and reproducibility, all developed test cases are made available in our reproduction package.

\section{Future Work}
While this study demonstrates comparison and generalization of employing agentic LLM systems for REST API test amplification, several directions remain open for future research.
First, our current workflows operate as linear pipelines without feedback from test execution or coverage results in the planning phase.
A natural next step is to integrate closed-loop feedback mechanisms, where structural coverage data from tools such as RESTCov~\cite{RESTCov_2025} can guide subsequent amplification iterations.
Such adaptive pipelines could enable self-improving testing agents that target under-tested API regions.

Second, although our evaluation covered multiple APIs of varying complexity, further studies on different industrial systems and LLM backends are needed to fully understand the scalability and portability of our approach.
This includes exploring open-weight models for reproducibility and assessing trade-offs between cost, performance, and energy efficiency.

Third, extending agent collaboration beyond the current planning and generation phases presents another promising research avenue.
For example, incorporating dedicated agents for oracle generation, fault localization, or test prioritization could lead to more comprehensive autonomous testing ecosystems.

Finally, future work should also investigate the human–AI collaboration dimension—how developers interact with agentic systems, validate their outputs, and integrate amplified test cases into continuous integration workflows.
Understanding these interactions is essential for ensuring the practical adoption of LLM-driven testing in real-world software engineering environments.

\section{Conclusion}
This paper investigates the capability of agentic LLM systems to amplify REST API test cases and provides a comparison across diverse cloud applications.
The study includes the evaluation of five different cloud applications, including open-source (PetStore, VAmPI, and Restful-Booker) and industrial-grade systems (Spotify and Google Drive).

\paragraph{RQ1: \RQTestStats} 
Many of the amplified test cases executed successfully and contributed substantially to the overall structural API coverage.
The amount of successful tests confirms that the generated tests are both syntactically valid and contextually aligned with the target APIs.
Additionally, the human effort in the refinement process is very minimal.
It shows that the amplification process produced executable, meaningful tests with minimal human intervention.

\paragraph{RQ2: \RQFailTypes}
The most failed amplified tests were caused by assertion errors.
These failures occurred when the generated test case executed correctly but produced an unexpected response compared to the expected output. 
A smaller number of failures were caused by runtime issues such as null pointer exceptions or other unhandled errors.
This pattern was consistent, including both open-source and industrial-grade applications.
It suggests that the amplified tests reached valid execution paths but encountered logical or semantic mismatches rather than structural errors.
These failures indicate that agentic systems are capable of generating realistic tests capable of exercising boundary conditions; however, they struggle with providing the correct expected values.

\paragraph{RQ3: \RQCoverage}
Both configurations achieved high structural API coverage, with multi-agent amplification offering broader request and parameter exploration.
Most endpoints and operations were fully explored.
The multi-agent configuration achieved broader coverage in request type and parameter dimensions.
It explored a wider range of input variations and payload structures. 
The single-agent configuration achieved more stable results in response-related metrics such as status and response type.
These results were consistent across open-source and industrial-grade systems.
The coverage patterns show that the agentic approach can systematically exercise REST APIs and adapt to different levels of system complexity.

\paragraph{RQ4: \RQCosts}
The results show that resource consumption increases with system complexity.
Larger systems, such as Google Drive and Spotify, required more time, tokens, and energy to complete the amplification process.
Smaller systems such as VAmPI, Restful-Booker, and PetStore consumed fewer resources in both configurations.
The multi-agent configuration used more tokens and computation time than the single-agent, resulting in higher monetary and energy costs.
The only exception was VAmPI, where token and energy usage decreased in the multi-agent configuration because input tokens are priced higher than output tokens.
These results confirm that multi-agent amplification offers broader exploration capabilities but at the cost of higher computational overhead.

\paragraph{RQ 5: \RQReadMaintain}
The amplified test cases maintained high readability and structural quality.
The code followed consistent formatting and naming conventions, and the structure aligned well with common REST Assured practices.
The readability checklist confirmed that the produced tests were easy to follow and suitable for integration into existing projects.
The maintainability of the amplified tests was also high.
The code was modular and concise, most importantly, compatible with the existing test suite and framework.
These results show that the amplification process can add additional tests that are not only functional but also clean and maintainable for long-term use.

Overall, the results confirm that agentic LLM systems can effectively amplify REST API test suites across diverse cloud environments.
The comparison between single-agent and multi-agent configurations highlights the trade-off between exploration depth and computational cost.
The findings demonstrate that LLM-based amplification can generate realistic, maintainable, and high-coverage tests that scale from open-source to industrial-grade systems.

\begin{acks}
This work is supported by the Research Foundation Flanders (FWO) via the BaseCamp Zero Project under Grant number S000323N.
\end{acks}

\bibliographystyle{ACM-Reference-Format}
\bibliography{Besjes2025tosem}

\appendix

\section{Prompts}
\subsection{Single-Agent}
\begin{lstlisting}[style=javacode, caption={Single-Agent Prompt}, label={lst:singleAgentPrompt}]
"You are an expert in writing REST API tests. Your job is to generate as much high-quality tests as possible. You'll be given an example testcase to know what the structure of the testcases should look like. Your goal is to maximize response and request type coverage, parameter coverage and parameter value coverage.

You have access to the following tools:
- OpenAPI documentation retriever: use this to get more information about an endpoint (starting with a '/') or a schema (starting with an uppercase).
- Local test executor: use this tool to locally execute the tests and check if the testing code compiles correctly.

Make sure to make the tests compliant with what is expected according to the OpenAPI specs.

Test the given endpoint from the input extensively with different parameters, parameter values, expected status codes, response types and request types, and make sure to cover as many edge cases as possible to detect potential bugs. You can only write tests for that single endpoint. Maximize the operation en status test coverage.

These schemas are available in the OpenAPI tool:
Order, Customer, Address, Category, User, Tag, Pet, ApiResponse

You can use the following function to interact with the HTTP client:
get(String path, Map<String, Object> params, Map<String, Object> headers)
get(String path, Map<String, Object> params, Map<String, Object> headers, Boolean authenticate)
post(String path, Map<String, Object> params, Map<String, Object> headers, Object data)
post(String path, Map<String, Object> params, Map<String, Object> headers, Object data, Boolean authenticate)
put(String path, Map<String, Object> params, Map<String, Object> headers, Object data)
put(String path, Map<String, Object> params, Map<String, Object> headers, Object data, Boolean authenticate)
delete(String path, Map<String, Object> params, Map<String, Object> headers, Object data, Boolean authenticate)

If you want to add the authorization headers to the request, you can set the 'authenticate' parameter to true.

You are NOT allowed to create new classes or extra functions. If you want to add payload to the request, you can use the built-in Java datastructures. Send payloads like JSON and XML as raw strings.

The following files are present:
emptyFile.png, jpegImage.jpeg, largeImage.jpg, pngImage.png, smallImage.png, unsupportedformat.txt

Use chain of thoughts to plan a solution to the problem and generate the test cases accordingly. Gather context, analyze, define the problem, plan a solution, and then execute the plan.

The output should be a valid Java test class that tests the given endpoint with different parameters, parameter values, expected status codes, response types and request types. Make sure to maximize the operation, status code, request/response type, parameter and parameter value test coverage. Only output a complete single test file, nothing more. Generate as much testcases as possible."
\end{lstlisting}

\subsection{Multi-Agent}
\begin{lstlisting}[style=javacode, caption={OpenAPI Extraction Agent Prompt}, label={lst:exctractionAgentPrompt}]
"Retrieve all relevant endpoints and schema for the endpoint '{state['endpoint_under_test']}' from the OpenAPI documentation. Make sure to get schemas recursively, as endpoints and schemas can reference other schemas. Query all the relevant endpoint and schema descriptions recursively. Query all referenced endpoints or schemas at once. An endpoint starts with a forward slash (/), and a schema or definition starts with an uppercase letter. Do not include the #/components/schemas/ or #/definitions/ prefix in the query.

When you are done, reply with 'DONE', nothing more."
\end{lstlisting}

\begin{lstlisting}[style=javacode, caption={Header Agent Prompt}, label={lst:headerAgentPrompt}]
"{state['openapi_references']}

Above are the OpenAPI references for the endpoint {state['endpoint_under_test']}.
Your goal is to search for header values that maximize:
- Status Code Coverage (triggering all documented status codes)
- Request Type Coverage (testing all documented request/content-type types)
- Response Type Coverage (triggering and testing all documented response/accept types)
Search for and document the following:
- Header values that trigger documented status codes for each HTTP method.
- All documented request/response types for each HTTP method.
- Header values that are expected to trigger documented status codes.

Output Format:
- All the headers found matching the description above, written in the format 'There should be a <METHOD> request with header <Header-Name> set to <header value>.'
Keep your answer concise and to the point. Use chain-of-thought reasoning. Only output the headers in the desired format."
\end{lstlisting}

\begin{lstlisting}[style=javacode, caption={Parameter Agent Prompt}, label={lst:parameterAgentPrompt}]
"{state['openapi_references']}

Above are the OpenAPI references for the endpoint {state['endpoint_under_test']}.
Your goal is to search for parameter selections that maximize:
- Parameter Coverage (testing all the documented parameters at least once)
- Status Code Coverage (triggering all documented status codes)
Search for and document the following:
- Identify all parameters (both required and optional) for each HTTP method.
- Identify parameter combinations that trigger documented status codes.

Output Format:
- All parameters and parameter combinations as described above in the following format: 'There should be a <METHOD> request with parameters <parameter name>, <parameter 2 name>, ...'. Make sure to include every required and optional parameter at least once for every HTTP method. For example make one request with all parameters.
Keep your answer concise and to the point. Use chain-of-thought reasoning. Only output the parameters in the desired format."
\end{lstlisting}

\begin{lstlisting}[style=javacode, caption={Value Agent Prompt}, label={lst:valueAgentPrompt}]
"{state['openapi_references']}

Above are the OpenAPI references for the endpoint {state['endpoint_under_test']}.
Your goal is to search for parameter values that maximize:
- Parameter Value Coverage (testing all documented values)
- Status Code Coverage (triggering all documented status codes)
Search for and document the following:
- All documented values for parameters.
- Identify values that trigger documented status codes.

Output Format:
- The parameter values that should be tested in the following format: 'There should be a <METHOD> request with parameter <parameter name> set to <value>.'
Keep your answer concise and to the point. Use chain-of-thought reasoning. Only output the values in the desired format.
\end{lstlisting}

\begin{lstlisting}[style=javacode, caption={Planner Agent Prompt}, label={lst:plannerAgentPrompt}]
"### OpenAPI References
{state['openapi_references']}
### Headers
{state['header_testcases']}
### Parameters
{state['parameter_testcases']}
### Values
{state['value_testcases']}

Available files:
emptyFile.png
jpegImage.jpeg
largeImage.jpg
pngImage.png
smallImage.png
unsupportedformat.txt

### Instructions for Test Plan Generation
Your task is to generate comprehensive test descriptions that maximize API test coverage by combining the requirements above. The final test plan should focus maximizing these coverages:

Request/Response Type Coverage: Ensure all documented request and response content types are tested for each HTTP method.
Status Code Coverage: Design tests to trigger every documented status code and intentionally combine headers, parameters, and values to expose undocumented status codes.
Parameter Coverage: Ensure all documented parameters (both required and optional) are included in at least one test per method.
Parameter Value Coverage: Test all valid parameter values at least once and explore edge cases.

Combine multiple requirements to achieve broader coverage with fewer test cases. Include both positive and negative scenarios. Every requirement should be reflected in the final test plan.

### Output Format
For each test case, describe:
- Endpoint and HTTP method
- Test goal
- Input details (headers, parameters, values)
- Expected outcome (status code, etc.)
Your goal is to create a precise and effective test plan that ensures maximum coverage while minimizing redundant tests. Every requirement provided by the other agents should be reflected in the final test plan. Focus on maximizing status code, request/response type, parameter, and parameter value coverage. Generate as much testcases as possible."
\end{lstlisting}

\begin{lstlisting}[style=javacode, caption={Test Writer Agent Prompt}, label={lst:writerAgentPrompt}]
"Here is the initial test case:
{testcase_content}
Turn the following testcases into one giant test class as above.

Test Descriptions:
{planner_agent_message}

You can use the following REST methods:
- get(String path, Map<String, Object> params, Map<String, Object> headers)
- get(String path, Map<String, Object> params, Map<String, Object> headers, Boolean authenticate)
- post(String path, Map<String, Object> params, Map<String, Object> headers, Object data)
- post(String path, Map<String, Object> params, Map<String, Object> headers, Object data, Boolean authenticate)
- put(String path, Map<String, Object> params, Map<String, Object> headers, Object data)
- put(String path, Map<String, Object> params, Map<String, Object> headers, Object data, boolean authenticate)
- delete(String path, Map<String, Object> params, Map<String, Object> headers, Object data)
- delete(String path, Map<String, Object> params, Map<String, Object> headers, Object data, Boolean authenticate)

If you want to add the authorization headers to the request, you can set the 'authenticate' parameter to true.
"
\end{lstlisting}

\begin{lstlisting}[style=javacode, caption={Test Executor Agent Prompt}, label={lst:executorAgentPrompt}]
"Analyzing the test execution output to identify any compilation errors or runtime exceptions.
Extract all information about the tests that crashed because of a **compilation error** or **runtime exception**.
Ignore tests with assertion failures unless they indicate syntax, exceptions or compilation problems.
If no compilation or runtime errors are found, reply with 'NO_COMPILATION_ERRORS'.
Execution Output:
{execution_output}

Output format:
- Give clear feedback on how to fix all the compilation errors."
\end{lstlisting}

\begin{lstlisting}[style=javacode, caption={Test Repair Agent Prompt}, label={lst:repairAgentPrompt}]
"Given the following test file:
{testfile}
And the feedback from the test executor agent:
{messages[-1].content}

Ensure there are no errors left."
\end{lstlisting}

\section{Test Cases}
\label{sec:testcases}
\subsection{Developer-written Test Files}

\begin{lstlisting}[style=javacode, caption={Developer-written Test File as System Input on Google Drive System}, label={lst:developerTestCaseGoogleDrive}]
import io.restassured.http.ContentType;
import io.restassured.response.Response;
import org.testng.Assert;
import org.testng.annotations.BeforeClass;
import org.testng.annotations.Test;
import static io.restassured.RestAssured.given;

public class GetFilesTest {

    private String accessToken;

    @BeforeClass
    public void mintAccessToken() {
        String clientId = System.getenv("GOOGLE_CLIENT_ID");
        String clientSecret = System.getenv("GOOGLE_CLIENT_SECRET");
        String refreshToken = System.getenv("GOOGLE_REFRESH_TOKEN");

        if (clientId == null || clientSecret == null || refreshToken == null) {
            throw new IllegalStateException(
                    "Missing env vars: GOOGLE_CLIENT_ID, GOOGLE_CLIENT_SECRET, GOOGLE_REFRESH_TOKEN"
            );
        }

        Response tokenResp =
                given()
                        .contentType(ContentType.URLENC)
                        .formParam("client_id", clientId)
                        .formParam("client_secret", clientSecret)
                        .formParam("refresh_token", refreshToken)
                        .formParam("grant_type", "refresh_token")
                        .when()
                        .post("https://oauth2.googleapis.com/token")
                        .then()
                        .statusCode(200)
                        .extract().response();

        accessToken = tokenResp.jsonPath().getString("access_token");
        Assert.assertNotNull(accessToken, "access_token should not be null");
    }

    @Test
    public void getFilesHappyPath() {
        Response resp =
                given()
                        .auth().oauth2(accessToken)
                        .queryParam("pageSize", 5)
                        .queryParam("fields", "files(id,name)")
                        .when()
                        .get("https://www.googleapis.com/drive/v3/files");

        // If this fails, print Google's error to diagnose quickly
        if (resp.statusCode() != 200) {
            System.out.println("HTTP " + resp.statusCode());
            System.out.println(resp.asPrettyString());
        }
        Assert.assertEquals(resp.statusCode(), 200, "Drive files.list should return 200");
        System.out.println("Files: " + resp.asPrettyString());
    }
}
\end{lstlisting}

\begin{lstlisting}[style=javacode, caption={Developer-written Test Files as System Input on Spotify System - I}, label={lst:developerTestCaseSpotify-I}]
import client.RestAssuredClient;
import io.restassured.response.Response;
import org.testng.Assert;
import org.testng.annotations.Test;
import java.util.HashMap;
import java.util.Map;

public class Auth extends RestAssuredClient {
/*
    curl -X POST "https://accounts.spotify.com/api/token" \
     -H "Content-Type: application/x-www-form-urlencoded" \
     -d "grant_type=client_credentials&client_id=your-client-id&client_secret=your-client-secret"

 */
    @Test
    public void getAccessToken(){
        String clientId = System.getenv("SPOTIFY_CLIENT_ID");
        String clientSecret = System.getenv("SPOTIFY_CLIENT_SECRET");

        Map<String, Object> params = new HashMap<>();
        params.put("grant_type", "client_credentials");
        params.put("client_id", clientId);
        params.put("client_secret", clientSecret);

        Map<String, Object> headers = new HashMap<>();
        headers.put("Content-Type", "application/x-www-form-urlencoded");

        Response response = postAuth("/token", params, headers, null);

        Assert.assertNotEquals(response.getStatusCode(), "200");
        System.out.println(response.getBody().asString());
    }
}
\end{lstlisting}

\begin{lstlisting}[style=javacode, caption={Developer-written Test Files as System Input on Spotify System - II}, label={lst:developerTestCaseSpotify-II}]
package services.Artists;

import client.RestAssuredClient;
import io.restassured.response.Response;
import org.testng.Assert;
import org.testng.annotations.Test;

public class GetArtists extends RestAssuredClient {

    @Test
    public void getArtistByIdHappyPath () {
        String artistId = "7dGJo4pcD2V6oG8kP0tJRR"; // Eminem

        Response response = get("/artists/" + artistId, null, null, null);
        Assert.assertEquals(response.getStatusCode(), 200);
    }
}
\end{lstlisting}

\subsection{Amplified Test Cases}
\begin{lstlisting}[style=javacode, caption={Amplified Test Script on Google Drive- I (Tests the creation of a file with default visibility)}, label={lst:testCreateFileDefaultVisibility}]
    @Test
    public void testCreateFileDefaultVisibility() {
        Response resp = post("https://www.googleapis.com/upload/drive/v3/files",
                Map.of("ignoreDefaultVisibility", false),
                Map.of("Authorization", "Bearer " + accessToken, "Content-Type", "application/octet-stream", "Accept", "application/json"),
                "<file_content_from_emptyFile.png>", true);
        Assert.assertEquals(resp.statusCode(), 200, "Should return 200 for creating file with default visibility");
    }

\end{lstlisting}

\begin{lstlisting}[style=javacode, caption={Amplified Test Script on Google Drive- II (Tests the maximum page size on the GET /drives request.)}, label={lst:testGetDrivesWithMaxPageSize}]
    @Test
    public void testGetDrivesWithMaxPageSize() {
        Response resp = get("/drives",
                Map.of("pageSize", 100),
                Map.of("Authorization", "Bearer " + accessToken), true);
        Assert.assertEquals(resp.statusCode(), 200, "Should return 200");
        System.out.println(resp.asPrettyString());
    }
\end{lstlisting}

\begin{lstlisting}[style=javacode, caption={Amplified Test Script on Spotify - I (Tests the "Search Artist" functionality on Spotify)}, label={lst:searchArtistHappyPath}]
    @Test
    public void searchArtistHappyPath() {
        Map<String, Object> headers = new HashMap<>();
        headers.put("Authorization", "Bearer " + validAccessToken);
        Map<String, Object> params = new HashMap<>();
        params.put("q", "remaster track:Doxy artist:Miles Davis");
        params.put("type", "artist");

        Response response = get("/search", params, headers, null);
        Assert.assertEquals(response.getStatusCode(), 200);
    }
\end{lstlisting}

\begin{lstlisting}[style=javacode, caption={Amplified Test Script on Spotify - II (Attempts to add an invalid URL to the Queue and expects an error code 403)}, label={lst:addInvalidUriToQueue}]
    @Test
    public void addInvalidUriToQueue() {
        Map<String, Object> headers = new HashMap<>();
        headers.put("Authorization", "Bearer " + validAccessToken);
        Map<String, Object> params = new HashMap<>();
        params.put("uri", invalidTrackUri);

        Response response = post("/me/player/queue", params, headers, null);
        Assert.assertEquals(response.getStatusCode(), 403);
    }
\end{lstlisting}

\subsection{Customized-Framework Used with Rest-Assured}
\begin{lstlisting}[style=javacode, caption={Amplified Test Script on Spotify - II (Attempts to add an invalid URL to the Queue and expects an error code 403)}, label={lst:framework}]
package client;
import io.restassured.RestAssured;
import io.restassured.response.Response;
import io.restassured.specification.RequestSpecification;
import java.util.Map;
import static client.APIConstant.BASE_URL;
import static client.APIConstant.BASE_URL_AUTH;
public abstract class RestAssuredClient {
    RequestSpecification httpRequest;

    public Response get(String path, Map<String, Object> params, Map<String, Object> headers, Object body) {
        RestAssured.urlEncodingEnabled = false;
        setHttpRequest(params, headers, body);
        Response response = httpRequest.get(path);
        response.then().log().body();
        return response;
    }
    public Response post(String path, Map<String, Object> params, Map<String, Object> headers, Object body) {
        RestAssured.urlEncodingEnabled = false;
        setHttpRequest(params, headers, body);
        Response response = httpRequest.post(path);
        response.then().log().body();
        return response;
    }
    public Response postAuth(String path, Map<String, Object> params, Map<String, Object> headers, Object body) {
        RestAssured.urlEncodingEnabled = false;
        setHttpRequestforAuth(params, headers, body);
        Response response = httpRequest.post(path);
        response.then().log().body();
        return response;
    }
    public Response put(String path, Map<String, Object> params, Map<String, Object> headers, Object body) {
        setHttpRequest(params, headers, body);
        Response response = httpRequest.put(path);
        response.then().log().body();
        return response;
    }
    public Response delete(String path, Map<String, Object> params, Map<String, Object> headers, Object body) {
        setHttpRequest(params, headers, body);
        Response response = httpRequest.delete(path);
        response.then().log().body();
        return response;
    }
    public Response patch(String path, Map<String, Object> params, Map<String, Object> headers, Object body) {
        setHttpRequest(params, headers, body);
        Response response = httpRequest.patch(path);
        response.then().log().body();
        return response;
    }
    protected void setHttpRequest(Map<String, Object> params, Map<String, Object> headers, Object body) {
        httpRequest = RestAssured.given().log().all(true).baseUri(BASE_URL);
        httpRequest.header("Authorization", "Bearer " + "<Access Token Here>");
        if (params != null) {
            httpRequest.queryParams(params);
        }
        if (headers != null) {
            httpRequest.headers(headers);
        }
        if (body != null) {
            httpRequest.body(body);
        }
    }
    protected void setHttpRequestforAuth(Map<String, Object> params, Map<String, Object> headers, Object body) {
        httpRequest = RestAssured.given().log().all(true).baseUri(BASE_URL_AUTH);
        httpRequest.header("Content-Type", "application/x-www-form-urlencoded");
        if (params != null) {
            httpRequest.formParams(params);
        }
        if (headers != null) {
            httpRequest.headers(headers);
        }
        if (body != null) {
            httpRequest.body(body);
        }
    }
}
\end{lstlisting}

\end{document}